\documentstyle[aps,psfig,subeqn]{revtex}  
\newcommand{\Rey}{\mbox{\it Re}}           
\newcommand{\Reycrit}{\mbox{\it Re$_c$}}   
\newcommand{\base}{{\bf U}}                
\newcommand{\perb}{{\bf u}'}               
\newcommand{\emode}{\tilde{\bf u}}         

\newcommand{\ie}{i.e.~}
\newcommand{\etc}{etc.~}
\newcommand{\UC}{{\bf U}_{\rm C}}
\begin{document}
\draft
\title{STABILITY ANALYSIS OF PERTURBED PLANE COUETTE FLOW}
\author{Dwight Barkley} 
\address{
Mathematics Institute, \\
University of Warwick, Coventry, CV4 7AL, United Kingdom \\
barkley@maths.warwick.ac.uk
}

\author{Laurette S. Tuckerman}
\address{
LIMSI-CNRS, BP 133, 91403 Orsay Cedex, France \\
laurette@limsi.fr
}
\date{\today} 
\maketitle 
 
\begin{abstract} 
Plane Couette flow perturbed by a spanwise oriented ribbon, similar to a
configuration investigated experimentally at the Centre d'Etudes de Saclay, 
is investigated numerically using a spectral-element code.  2D steady 
states are computed for the perturbed configuration; these differ from 
the unperturbed flows mainly by a region of counter-circulation
surrounding the ribbon.  The 2D steady flow loses stability to 3D
eigenmodes at $\Rey_c = 230, \beta_c = 1.3$ for $\rho = 0.086$ and
$\Rey_c \approx 550, \beta_c \approx 1.5$ for $\rho = 0.043$, where
$\beta$ is the spanwise wavenumber and $2\rho$ is the height of the
ribbon.  For $\rho = 0.086$, the bifurcation is determined to be 
subcritical by calculating the cubic term in the normal form equation 
from the timeseries of a single nonlinear simulation; 
steady 3D flows are found for $\Rey$ as low as 200.
The critical eigenmode and nonlinear 3D states contain streamwise 
vortices localized near the ribbon, whose streamwise extent increases
with $\Rey$. 
All of these results agree well with experimental observations.
\end{abstract} 


\section{INTRODUCTION}

It is well known that, of the three shear flows most commonly used to
model transition to turbulence, plane Poiseuille flow is linearly
unstable for $\Rey > 5772$, whereas pipe Poiseuille flow and 
plane Couette flow are linearly stable for all Reynolds numbers;
see, e.g. \cite{Bayly}.
Yet, as is also well established, in laboratory experiments, plane
and pipe Poiseuille flows actually undergo transition to three-dimensional 
turbulence for Reynolds numbers on the order of $1000$.
For plane Couette flow, the lowest Reynolds numbers at which
turbulence can be produced and sustained has been shown to be
between 300 and 400 both in numerical simulations \cite{Lundbladh,Hamilton}
and in experiments \cite{Tillmark,Daviaud}.

The gap between steady, linearly stable flows which depend on only one
spatial coordinate
and three-dimensional turbulence can be bridged by studying perturbed
versions of Couette and Poiseuille flow.  
Plane Couette flow perturbed by a wire midway between
the bounding plates and oriented in the spanwise direction has been
the subject of laboratory experiments by Dauchot and
co-workers \cite{Dauchot,Bottin,Manneville} at CEA-Saclay.
Our goal in this paper is to study numerically the flows and 
transitions in a configuration similar to that of the Saclay
experiments.

Previous studies of plane channel flows
have used a variety of approaches.
We briefly review these, emphasizing 
computational investigations and the plane Couette case.

One approach is to seek finite amplitude solutions at transition
Reynolds numbers and to understand the dynamics of transition in terms
of these solutions.  Finite amplitude solutions for plane Couette flow
have been found for Reynolds numbers as low as $\Rey=125$ by 
numerically continuing steady states
or travelling waves from other flows: the wavy Taylor vortices of
cylindrical Taylor-Couette flow by Nagata \cite{Nagata,Nagata98} and 
Conley and Keller \cite{Keller} and the 
wavy rolls of Rayleigh-B\'enard convection by Busse \cite{Busse}.  Most
recently, Cherhabili and Ehrenstein \cite{Cherhabili95,Cherhabili97}
succeeded in continuing
plane-Poiseuille-flow solutions to plane Couette flow, via an
intermediate Poiseuille-Couette family of flows.  They showed that 
in proceeding from Poiseuille to Couette flow,
the wavespeed of the travelling waves decreases and their streamwise
wavelength increases, as does the number of harmonics needed to
capture them. When the Couette limit is reached, the finite amplitude
solutions are highly streamwise-localized steady states.  
The minimum Reynolds number achieved in these continuations is $\Rey=1500$.
None of these steady solutions of plane Couette flow obtained so far
are stable.

A second, highly successful, approach has been to study the transient
evolution of linearized plane Couette flow.  Although all initial
conditions must eventually decay and the most slowly decaying mode
must be spanwise invariant by Squire's theorem, the non-normality of
the evolution operator allows large transient growth.  
Butler and Farrell \cite{Butler} showed that a thousand-fold growth
in energy could be achieved from an initial condition resembling streamwise 
vortices which are approximately circular and streamwise invariant.
Reddy and Henningson \cite{Reddy} computed the maximum achievable growth 
for a large range of Reynolds numbers.
An interpretation is given by these authors and by Trefethen et 
al. \cite{Trefethen} in terms of pseudospectra: the spectra of non-normal
operators display an extreme sensitivity to perturbations of the
operator.  Thus, slightly perturbed plane Couette or Poiseuille flows
may be linearly unstable for much lower Reynolds numbers than the
unperturbed versions.

A third broad category of computational investigation is the study of
nonlinear temporal evolution in relatively tame turbulent plane
channel flows.  Orszag and Kells \cite{Kells} and 
Orszag and Patera \cite{Patera}
showed that finite amplitude spanwise-invariant states of plane
Poiseuille flow are unstable to 3D perturbations; this is also true of
quasi-equilibria for plane Poiseuille and Couette flow.  
Lundbladh and Johansson \cite{Lundbladh} showed that turbulent
spots evolved from initial disturbances resembling streamwise
vortices if the Reynolds number exceeded a critical Reynolds
number between 350 and 375.
Numerical simulations by Hamilton, Kim and Waleffe \cite{Hamilton} 
of turbulent plane Couette flow at $Re=400$
indicated that streamwise vortices and streaks played an important role 
in a quasi-cyclic regeneration process.  
Coughlin \cite{Coughlin} used weak forcing to stabilize steady states 
containing streamwise vortices and streaks.
These became unstable and underwent a similar regeneration cycle 
when the forcing or Reynolds number was increased.
The critical Reynolds numbers displayed in all of these numerical 
simulations are in good agreement with experiments
by Tillmark and Alfredsson \cite{Tillmark} 
and by Daviaud et al. \cite{Daviaud} who reported turbulence at
$\Rey \gtrsim 360$ and $\Rey \gtrsim 370$, respectively.

The last approach we discuss, and the most relevant to this study,
is perturbation of the basic shear profile,
to elicit instabilities that are in some sense nearby.
If a geometric perturbation breaks either the
streamwise or spanwise invariance of the basic profile, 
then the flow is freed from the constraint of Squire's theorem, 
which would otherwise imply that the linear instability at lowest 
Reynolds number is to a spanwise invariant (2D) eigenmode.  
A perturbed flow with broken
symmetry may directly undergo a 3D linear instability.  
One can hope to understand the behavior of the unperturbed system 
by considering the limit in which the perturbation goes to zero.
For some time, experimentalists \cite{Nishioka} have used 
perturbations to produce spanwise-invariant Tollmien-Schlichting waves 
arising subcritically.
More recently, for example, Schatz et al. \cite{Schatz} inserted 
a periodic array of cylinders in a plane Poiseuille experiment 
to render this bifurcation supercritical.
In plane Couette flow, Dauchot and co-workers at Saclay
\cite{Dauchot,Bottin,Manneville} found that streamwise
vortices could be induced for Reynolds numbers around 200 when a wire
was placed in the flow (the exact range in Reynolds number for
which the vortices occur depends on the radius of the wire).  
They suspected that these vortices arise from a subcritical bifurcation 
from the perturbed profile, but did not determine this.   

\begin{figure}[h]
\vspace*{-4cm} \centerline{\psfig{file=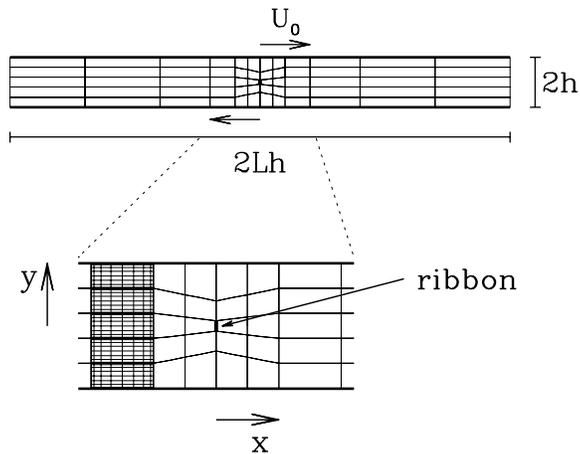,width=15cm}}
\vspace*{-4.5cm}
\caption{{} Flow geometry considered in the paper.  The upper and
lower channel walls are separated by distance $2h$ and move with
velocities $U_0\hat x$ and $-U_0\hat x$, repectively.  An infinitely
thin ribbon (bold line) is located midgap in the channel and has
height $2\rho h$ ($\rho=0.086$ for the case shown).  The
computational mesh (macro elements) used in our calculations is shown,
as is the (fine) collocation mesh for polynomial order $N=8$ (three
elements in the enlargement).  The ribbon is formed by setting no-slip
boundary conditions on the edges of two adjoining elements.  Periodic
boundary conditions are imposed over length $2Lh$ in the horizontal
direction.  The full geometry shown has aspect ratio $L=10$.  The
system is homogeneous in the spanwise ($z$)-direction normal to the
figure. }
\label{fig:mesh}
\end{figure}

In this paper, we numerically study the destabilization of plane
Couette flow when a ribbon is placed midway in the channel gap
(Fig.~\ref{fig:mesh}).  The ribbon is infinitely thin in the
streamwise ($x$) direction, occupies a fraction $\rho$ of the cross-channel 
($y$) direction, and is infinite in the spanwise ($z$) direction.
This geometry is similar, though not identical, to that used in the
Saclay experiments.  In the experiments, the perturbation is 
a thin wire with cylindrical cross-section.  
Here we use a ribbon because it is much easier to simulate numerically.
For the experimental or numerical results to be of wider importance, 
the particular shape of the perturbation should
not be important, as long as it is small.

We shall address the extent to which a small geometric perturbation
of the plane Couette geometry affects the stability of the flow.
We will show that a small geometric perturbation does indeed lead to a
subcritical bifurcation to streamwise vortices,
at Reynolds numbers and wavenumbers which agree well with the Saclay experiments.

\section{NUMERICAL COMPUTATIONS}

The computations consist of three parts: 
(1) obtaining steady 2D solutions of the Navier--Stokes equations, 
(2) determining the linear stability of these solutions to 3D 
perturbations, and
(3) classifying the bifurcation via a nonlinear stability analysis.
Here we outline the numerical techniques for carrying out 
these computations.

\subsection{2D steady flows}
\label{sec:num2D}

Our computational domain has been shown in fig. \ref{fig:mesh}.
We non-dimensionalize lengths by the channel half-height $h$,
velocities by the speed $U_0$ of the upper channel wall, time by the
convective time $h/U_0$.
There are two nondimensional parameters for the flow,
which we take to be the usual Reynolds number for plane
Couette flow, $\Rey= h U_0/\nu$, where $\nu$ is the kinematic
viscosity of the fluid,
and the nondimensional half-height of the ribbon $\rho$,
hereafter called its radius for consistency with the Saclay experiments.
We view the (nondimensionalized) streamwise periodicity 
length $2L$ as a numerical parameter which we take sufficiently 
large that the system
behaves as though it were infinite in the streamwise direction.

The fluid flow is governed by the incompressible Navier--Stokes
equations:
\begin{subequations}
\begin{eqnarray}
 &{\displaystyle\frac {{\partial {\bf u}}}{\partial t}} = 
   -({\bf u} \cdot \nabla){\bf u} - \nabla p 
   + {\displaystyle\frac{1}{Re}} \nabla^2 {\bf u} & ~~~\hbox{in $\Omega$}, \label{nse}\\
 &\nabla \cdot {\bf u} = 0 & ~~~\hbox{in $\Omega$},
\label{divcond}
\end{eqnarray}
\label{nsedivcond}
\end{subequations}
subject to the boundary conditions:
\begin{subequations}
\begin{eqnarray}
&&{\bf u}(x-L,y) = {\bf u}(x+L,y) \label{bcs:periodic} \\
&&{\bf u}(x,y=\pm 1)= \pm {\bf \hat x} \label{bcs:Couette} \\
&&{\bf u}(x=0,y)=0, ~~~~ \quad\hbox{for $-\rho \le y \le \rho$}, \label{bcs:ribbon}
\end{eqnarray}
\label{bcs}
\end{subequations}
where ${\bf u}\equiv(u,v,w)$ is the velocity field, $p$ is the 
nondimensionalized static pressure and $\Omega$ is the
computational domain.  
The pressure $p$, like ${\bf u}$, satisfies periodic
boundary conditions in $x$.

Time-dependent simulations of these equations in two dimensions
($w\equiv 0$, $\partial/\partial z \equiv 0$) are carried out using the
spectral element \cite{Patera84} program {\em Prism} \cite{Henderson94,Karniadakis}.
In the spectral element method, the domain is
represented by a mesh of macro elements as shown in
Fig.~\ref{fig:mesh}.  The channel height is spanned by five elements
while the number of elements spanning the streamwise direction depends
on its length: 24 elements are used for $L=32$ and $36$ elements for
$L=56$.  The no-slip condition (\ref{bcs:ribbon}) is enforced by
setting zero velocity boundary conditions along the edges of two
adjoining mesh elements: this interface defines the ribbon.  If
continuity were imposed along this interface, 
as is done on all other element boundaries,
then the flow would reduce to unperturbed plane Couette flow.  Thus
the ribbon is modeled by a small (but significant) change in the
boundary conditions on just two edges of elements in the computational
domain.  Within each element both the geometry and the solution
variables (velocity and pressure) are represented using $N$th order
tensor-product polynomial expansions.  The collocation mesh in
Fig.~\ref{fig:mesh} (enlargement) corresponds to an expansion with $N=8$.

A time-splitting scheme is used to integrate the underlying
discretized equations \cite{Israeli}.
Based on simulations with polynomial order $N$ in the range $6 \le N
\le 12$ and timesteps $\triangle t$ in the range $10^{-3} \le
\triangle t \le 10^{-2}$ we have determined that $N=8$ and $\triangle
t=0.005$ give valid results over the range of $\Rey$ considered.
These numerical parameter values (typical for studies of this type)
have been used for most of the results reported. Each velocity
component is thus represented by about 7500 scalars for $L=32$.

Steady flows used for our stability calculations have been obtained
from simulations with Reynolds numbers in the range $100 \leq Re \leq
600$.  In all cases, the simulations were run sufficiently long to
obtain asymptotic, steady velocity fields.  We shall denote these
steady 2D flows by $\base(x,y)$.

\subsection{Linear stability analysis}
\label{sec:numlin}

Let $\base(x,y)$ be the 2D base flow whose stability is sought.  An
infinitesimal three-dimensional perturbation $\perb(x,y,z,t)$ evolves
according to the Navier--Stokes equations linearized about $\base$.
Because the resulting linear system is homogeneous in the spanwise
direction $z$, generic perturbations can be decomposed into Fourier
modes with spanwise wavenumbers $\beta$:
\begin{equation}
\label{bmode}
\begin{array}{rcl}
{\bf \perb}(x,y,z,t) &=& (\hat{u} \cos\beta z, \,
                          \hat{v} \cos\beta z, \,
                          \hat{w} \sin\beta z) \\ 
         p'(x,y,z,t) &=&  \hat{p} \cos\beta z 
\end{array}
\end{equation}
or an equivalent form obtained by translation in $z$.  
The vector ${\bf \hat u} (x,y,t)= ({\hat u},{\hat v},{\hat w})$ of Fourier
coefficients evolves according to:
\begin{subequations}
\label{nsedivcondlin}
\begin{eqnarray}
 &{\displaystyle\frac{\partial {\bf \hat u}}{\partial t}} 
   =  -({\bf \hat u} \cdot \nabla){\base}
   -({\bf \base} \cdot \nabla){\bf \hat u} 
   - (\nabla -\beta {\bf \hat z}) {\hat p }
   + {\displaystyle\frac{1}{Re}} (\nabla^2 -\beta^2) {\bf \hat u} & ~~~\hbox{in $\Omega$}, 
\label{nselin}\\
&(\nabla + \beta {\bf \hat z}) \cdot {\bf \hat u} =  0 & ~~~\hbox{in $\Omega$},
\label{divcondlin}
\end{eqnarray}
\end{subequations}
where $\nabla$, \etc are two-dimensional differential operators.
Equations (\ref{nsedivcondlin}) are solved subject to homogeneous boundary conditions:
\begin{subequations}
\begin{eqnarray}
{\bf \hat u}(x-L,y) &=& {\bf \hat u}(x+L,y) \label{bcslin:periodic} \\
{\bf \hat u}(x,y=\pm 1) &=& 0 \label{bcslin:Couette} \\
{\bf \hat u}(x=0,y) &=& 0, ~~~~ \quad\hbox{for $-\rho \le y \le \rho$}, 
\label{bcslin:ribbon}
\end{eqnarray}
\label{bcslin}
\end{subequations}
Equations (\ref{nsedivcondlin}) with boundary conditions
(\ref{bcslin}) can be integrated numerically by the method described
in section \ref{sec:num2D}.  For fixed $\beta$, this is essentially a
two-dimensional calculation \cite{Barkley,Henderson96}.
After integrating (\ref{nsedivcondlin})-(\ref{bcslin}) a sufficiently
long time, only eigenmodes corresponding to leading eigenvalues
remain.  We use this to find the leading eigenvalues (those with
largest real part) and corresponding eigenmodes for fixed values of
$\Rey$ and $\beta$ as follows.
A Krylov space is constructed based on integrating
(\ref{nsedivcondlin})-(\ref{bcslin}) over $K=8$ successive
(dimensionless) time intervals of $T=5$.
More precisely, we calculate the fields 
${\bf \hat u}(t), {\bf \hat u}(t+T), \ldots {\bf \hat u}(t+(K-1)T)$
and orthonormalize these to form a basis $v_1, v_2, \ldots v_K$.
We then define the $K \times K$ matrix 
$H_{ij} \equiv \langle v_i, {\bf L} v_j \rangle$
where ${\bf L}$ is the operator on the right-hand-side
of the linearized Navier-Stokes equations
and $\langle,\rangle$ is an inner product.
Approximate eigenvalues $\sigma$ and eigenmodes $\emode(x,y,z)$ 
are calculated by diagonalizing $H$; 
their accuracy is tested by computing the residual 
$r \equiv || \sigma {\bf \tilde u}- {\bf L \tilde u} ||$.
If the eigenvalue-eigenmode pairs do not attain a desired accuracy ($r <
10^{-5}$ for the case here), then another iteration is performed. The
new vector is added to the Krylov space and the oldest vector is
discarded.  This is effectively subspace iteration initiated with a
Krylov subspace.  More details can be found in \cite{Schatz,Barkley,Mamun}.

\begin{figure}[h]
\vspace*{-5cm}
\centerline{\psfig{file=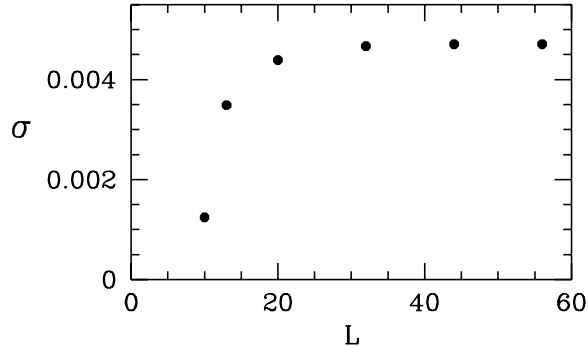,width=12cm}}
\vspace*{-2cm}
\caption{{}Leading eigenvalue as a function of streamwise periodicity
half-length $L$ for $\Rey=250$, $\beta=1.3$.  
For $L \protect\gtrsim 32$ the eigenvalue is independent of $L$.}
\label{fig:ev_L}
\end{figure}

We conclude this section by considering the effect of the streamwise
periodicity length $2L$ on the computations.  Recall that we view $L$
as a quasi-numerical parameter in that we seek solutions valid for
large $L$.  Figure~\ref{fig:ev_L} shows the dependence of the leading
eigenvalue $\sigma$ on streamwise length at $\Rey=250$, $\beta=1.3$
(values near the primary 3D linear instability).  It can be seen that
for $L \gtrsim 32$ the eigenvalue is independent of $L$.  This is
consistent with the structure of the base flow and eigenmodes shown in
the following section.  Most of the computations reported have used
$L=32$.

\subsection{3D simulations}
For the nonlinear stability analysis and for obtaining steady 3D flows, 
we carry out 3D simulations of
(\ref{nsedivcond})-(\ref{bcs}) using the same spectral element
representation in $(x,y)$ described above together with a Fourier
representation in the spanwise direction $z$.  We impose periodicity
in the spanwise direction by including wavenumbers $m\beta_c$ for
integers $|m| < M/2$, where $\beta_c$ is the critical wavenumber found
in the linear stability analysis.  
The simulations we report use $M=16$.

\section{Results}

\subsection{2D Steady flows}
\label{sec:steady}

\begin{figure}[ht]
\vspace*{-4cm}
\centerline{\psfig{file=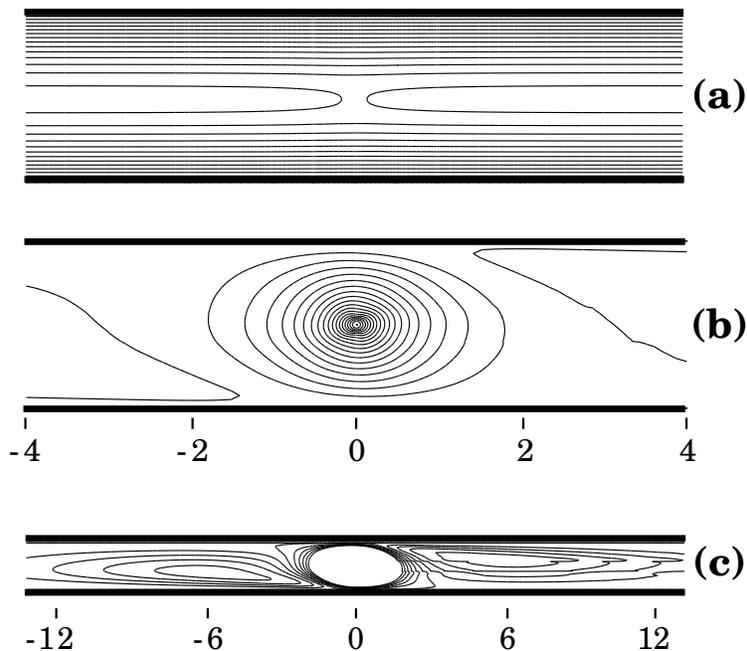,width=10cm}}
\vspace*{-3cm}
\caption{{} The steady two-dimensional base flow $\base(x,y)$ at
$\Rey=250$ for a ribbon with $\rho=0.086$.  Only the central portion
of the full $L=32$ domain is shown. (a) Streamfunction contours of
$\base$.  The flow is nearly identical to the parallel shear of plane
Couette flow except very near the ribbon.  (b) Streamfunction contours
of the deviation $\base - \UC$ highlighting the difference between the
perturbed and unperturbed Couette flows.  A region ($|x| \protect\lesssim 3$) 
of positive circulation is established around the ribbon.  The flow is
centro-symmetric. (c) The deviation over a larger streamwise extent
showing regions ($3 \protect\lesssim |x| \protect\lesssim 24$) 
further from the ribbon whose circulation is negative,
like that of $\UC$.  The flow is very weak; contours of the dominant
part of this flow are not shown. The slight lack of centro-symmetry is
a graphical artifact.}
\label{fig:base}
\end{figure}

A typical steady 2D flow for the perturbed Couette geometry is shown
in Fig.~\ref{fig:base}.  It is representative of base flows for
Reynolds numbers on the order of a few hundred with a ribbon of size 
$\rho=0.086$.  This was chosen to correspond to the radius of 
one of the cylinders used in the Saclay 
experiments \cite{Dauchot,Bottin,Manneville}.  The Reynolds
number $\Rey = 250$ of the flow shown is close to the threshold for
the 3D instability that will be discussed in the next section.
Unless otherwise stated, results are for $\rho = 0.086$, $\Rey = 250$,
and $L = 32$.

In Fig.~\ref{fig:base}(a) it can been seen that, except near the
ribbon, the steady flow is essentially the parallel shear of
unperturbed plane Couette flow.  The streamlines are as reported
experimentally in \cite{Dauchot}.  As noted there, the Reynolds
number based on the radius of the ribbon and the local velocity near
the ribbon is very small compared to Reynolds numbers where separation
or vortex shedding could be expected.  In Fig.~\ref{fig:base}(b) we
plot the streamfunction of the deviation $\base - \UC$ where $\UC =
y{\bf \hat x}$ is the unperturbed plane Couette profile.  The primary
effect of the ribbon is to establish a region ($|x| \lesssim 3 $) of
positive circulation (opposing that of plane Couette flow) surrounding
the ribbon.  Figure \ref{fig:base}(c) shows $\base - \UC$ over a
larger streamwise extent.  Further from the ribbon are wider regions
($3 \lesssim |x| \lesssim 24$) in which the deviation is weak, but has
the same negative circulation as plane Couette flow.

The size of the counter-rotating region is remarkably uniform over the
ribbon radii $\rho$ and Reynolds numbers $Re$ that we have studied.
We define the streamwise extent of the counter-rotating region as
delimited by $\psi(x,y=0) = 0$, i.e. the $x$ values at which the
streamfunction at midheight $y=0$ has the same value as at the channel
walls $y=\pm 1$.  For $\rho = 0.086$, the counter-rotating region
varies from $|x|\leq 2.18$ for $Re = 150$ to $|x|\leq 3.00$ for $Re =
300$, while for $\rho = 0.043$ the counter-rotating region varies from
$|x| \leq 2.10$ for $Re = 150$ to $|x| \leq 2.87$ for $Re = 600$.
This insensitivity to the size of $\rho$ is significant in light of
the 2D finite-amplitude steady states calculated by Cherhabili and
Ehrenstein \cite{Cherhabili95,Cherhabili97}.  
The states found by these authors in unperturbed
plane Couette flow strongly resemble that in Fig.~\ref{fig:base}.
These too have a central counter-rotating region surrounded by larger
regions of negative circulation.  At $Re = 2200$, the counter-rotating
region in their flow occupies $|x| \leq 2.31$ 
(see Figs.~10 and 11 of \cite{Cherhabili95}, 
Figs.~2 and 3 of \cite{Cherhabili97})
The similarity between the 2D flows for $\rho = 0.086$, $\rho = 0.043$, 
and, effectively, $\rho = 0$ leads us to hypothesize that our 2D perturbed
plane Couette flows are connected (via the limit $\rho \rightarrow 0$)
to those computed by Cherhabili and Ehrenstein.

We may also quantify the intensity of the counter-circulation.  One
measure is the maximum absolute value of $v$, which is attained very
near the ribbon, at $(x,y) = (\pm0.081,0)$.  This value is
approximately independent of Reynolds number, but decreases strongly
with ribbon radius: $v_{max}\approx 0.031$ for $\rho = 0.086$ and
$v_{max}\approx 0.013$ for $\rho = 0.043$.

An important qualitative feature of the flow can been seen in
Figs.~\ref{fig:base}(b) and (c): the flow is centro-symmetric, i.e. it
is invariant under combined reflection in $x$ and $y$, or equivalently
rotation by angle $\pi$ about the origin.  It can be verified that the
governing equations (\ref{nsedivcond}) and boundary conditions
(\ref{bcs}) are preserved by the centro-symmetric transformation:
\begin{equation}
{\bf u}(x,y) \rightarrow -{\bf u}(-x,-y).
\label{centro}
\end{equation}
The unperturbed plane Couette problem is also centro-symmetric.  It is
in fact symmetric under the Euclidean group $E_1$ of translations and
the ``reflection'' consisting of the centro-symmetric transformation
(\ref{centro}).  The ribbon in the perturbed flow breaks the
translation symmetry, but leaves the centro-symmetry intact.  Note
that reflections in $x$ or $y$ alone are not symmetries of either the
unperturbed or the perturbed plane Couette problem because either
reflection alone reverses the direction of the channel walls,
violating the boundary conditions (\ref{bcs:Couette}).

\begin{figure}[h]
\vspace*{-3cm}
\centerline{\psfig{file=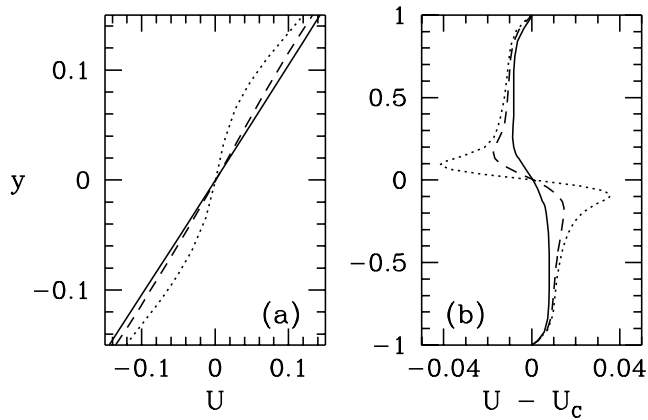,width=12cm}}
\vspace*{-1cm}
\caption{{} Streamwise velocity profiles in the perturbed geometry.
(a) $U(x,y)$ as a function of $y$ for $x = 0.081$ (dotted), $x = 0.25$ (dashed),
and $x = 0.5$ (solid).  Only the central portion of the channel is
shown.  Only very close to the ribbon does the velocity differ
significantly from the linear profile.  (b) Deviation $U(x,y)-y$ over
the full range of $y$.}
\label{fig:base_profiles}
\end{figure}

In Fig.~\ref{fig:base_profiles} we present streamwise velocity
profiles near the ribbon.  For $|x| > 0.5$, the Couette profile is very 
nearly recovered.  Figure~\ref{fig:base_profiles}(b) shows streamwise
velocity profiles of the deviation from the linear Couette profile
across the full channel.  Close examination reveals that these
profiles are not odd in $y$, consistent with the fact that the system
is neither symmetric nor antisymmetric under reflection in $y$.  The
symmetric partners to the profiles shown are at negative $x$ values.

The profiles in Fig.~\ref{fig:base_profiles} are similar to those
Bottin et al. \cite{Manneville} 
obtained in the Saclay experiments under similar conditions.
It is not possible to compare directly with experiment because of the
difficulty in obtaining experimental velocity profiles and because the
geometric perturbations differ in the computations and experiments.
The only noticeable difference between experiments and computations is
that the profiles Fig.~\ref{fig:base_profiles}(b) are very nearly odd
in $y$, whereas in experiment this lack of symmetry is more
pronounced.

\begin{figure}[h]
\vspace*{-3.5cm}
\centerline{\psfig{file=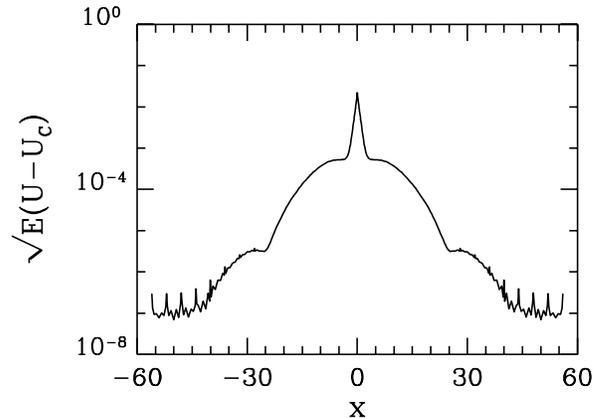,width=12cm}}
\vspace*{-1.5cm}
\caption{ {} Energy of deviation between perturbed and unperturbed 
Couette flows as a function of $x$.
Parameters are the same as in Fig.~\protect{\ref{fig:base}} 
except that here $L=56$.  
Abrupt changes in slope at $|x| = 2.76, |x| = 23.78$
correspond to changes in the sign of the circulation of 
$\base - \UC$.
For $|x| \protect\gtrsim 40$,
the deviation is below the precision of the computations.}
\label{fig:enbmc}
\end{figure}

Finally in Fig.~\ref{fig:enbmc} we quantify the deviation between
perturbed and unperturbed plane Couette flow by plotting the energy
per unit length $E(\base-\UC) \equiv \int_{-1}^1 dy {1\over 2}|\base(x,y)-\UC(y)|^2$ 
as a function of $x$ for $-56 \le x \le 56$.  The data show a narrow
central region, corresponding to the region $|x| \leq 2.76$ of
positive circulation seen in Fig.~\ref{fig:base}(b), where the
deviation falls sharply and approximately exponentially in $x$.  For
$|x|>2.76$, the deviation, while very small, decays very slowly (and not
exponentially) with $|x|$.  The boundaries $x = \pm23.78$ terminating
the outer region of negative circulation can also be seen on
fig.~\ref{fig:enbmc}.  The precision of the computations is surpassed
beyond $|x|=40$.  This figure shows that for $|x|>32$ the deviation of
the base flow from Couette is indeed very weak and this supports our
choice of $L=32$ as an adequate domain size for most computations.

\subsection{3D Linear stability results}
\label{sec:linear}

The two-dimensional steady flows just discussed become linearly
unstable to three-dimensional perturbations when the Reynolds number
exceeds a critical value $Re_c$.  To determine this value and the
associated wavenumber, we have performed a linear stability analysis
of the steady flows via the procedure described in
Sec.~\ref{sec:numlin}.

\begin{figure}[ht]
\vspace*{-3cm}
\centerline{\psfig{file=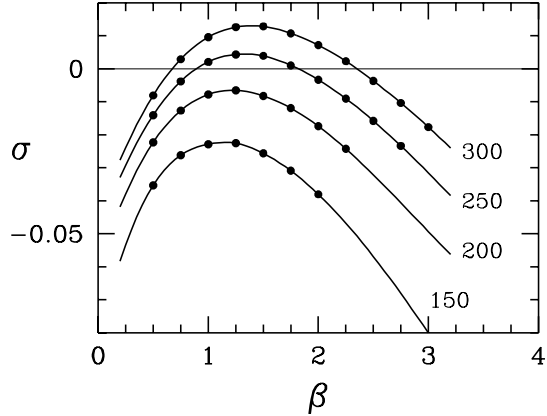,width=12cm}}
\vspace*{-1cm}
\caption{{}  Growth rate $\sigma$ of most unstable three-dimensional
eigenmode as a function of spanwise wavenumber $\beta$ for
$Re=150,200,250,300$ with ribbon radius $\rho = 0.086$.  Critical
values for instability are $Re_c=230$ and $\beta_c=1.3$. }
\label{fig:sigma86}
\end{figure}

Figure~\ref{fig:sigma86} shows the growth rate $\sigma$ of the most
unstable three-dimensional eigenmode $\emode$ as a function of $\Rey$ and
spanwise wavenumber $\beta$ for a ribbon with $\rho=0.086$.  For each
value of $\Rey$, we have fit a piecewise-cubic curve, shown in
fig.~\ref{fig:sigma86}, through the eigenvalue data to determine the
wavenumber $\beta_{\rm max}(\Rey)$ which maximizes $\sigma$.  The
critical Reynolds number $\Rey_c$ is then determined by linear
interpolation of $\sigma(\beta_{\rm max}(\Rey),\Rey)$ through these
maxima and finding its zero crossing.  From this we find critical
values for the onset of linear instability to be $\Rey_c=230$ and
$\beta_c=1.3$ for the ribbon with $\rho = 0.086$.  These values are
consistent with what is seen experimentally, but we delay discussion
until the conclusion.

\begin{figure}[ht]
\vspace*{-4cm}
\centerline{\psfig{file=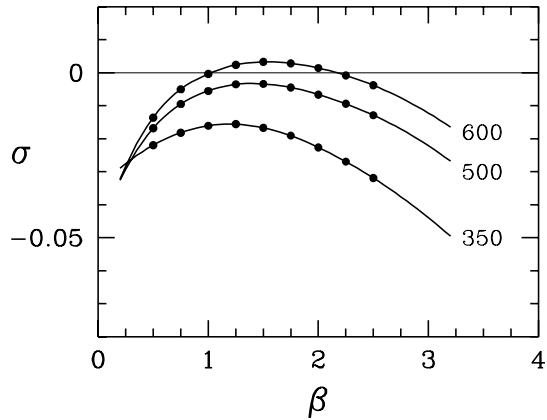,width=12cm}}
\vspace*{-1cm}
\caption{{}Growth rate of most unstable eigenmode 
as a function of $\beta$ and $\Rey$ for $\rho = 0.043$.
Critical values are $Re_c \approx 550$ and $\beta_c \approx 1.5$.}
\label{fig:sigma43}
\end{figure}

Figure~\ref{fig:sigma43} shows similar eigenvalue spectra for a ribbon
half as large: $\rho = 0.043$.  The critical wavenumber $\beta_c
\approx 1.5$ is only slightly larger than the previous value.  However, the
critical Reynolds number is much larger: $\Rey_c \approx 550$.  The critical
Reynolds number must increase as $\rho$ is decreased since, when no
ribbon is present, the problem reduces to classical plane Couette flow
which is linearly stable for all finite $\Rey$, \ie $\lim_{\rho
\rightarrow 0} Re_{\rm c}(\rho) = \infty$.

We note that Cherhabili and Ehrenstein \cite{Cherhabili97} also 
calculate 3D instability for their 2D finite amplitude plane Couette flows.
Despite the
resemblance of their 2D flows to ours, the spanwise wavenumber
corresponding to maximal growth is much larger in their case: $\beta
\approx 23$.

\begin{figure}[ht]
\vspace*{-3cm}
\centerline{\psfig{file=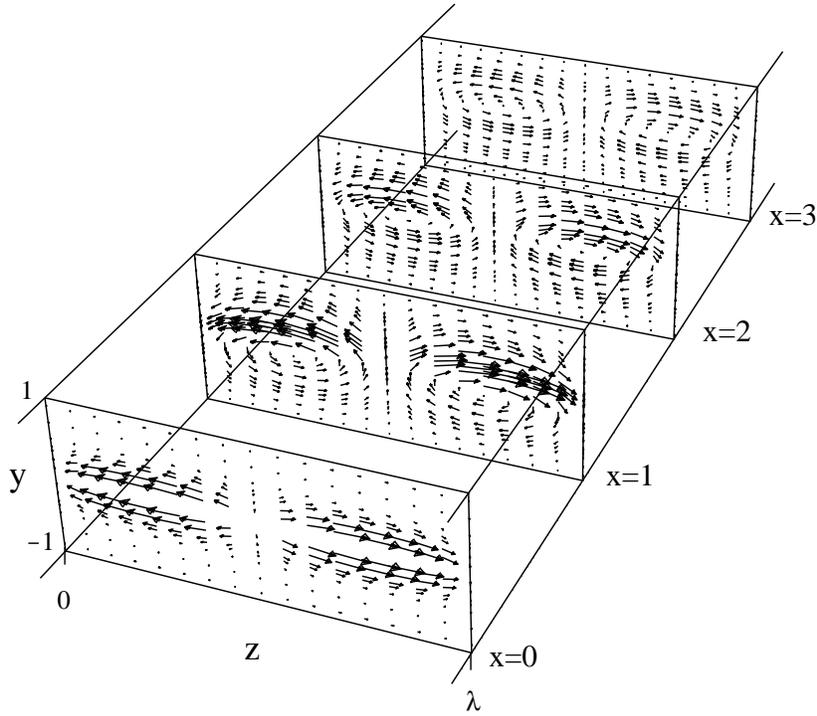,width=12cm}}
\vspace*{-2cm}
\caption{{}Velocity field of near-marginal eigenvector in the planes $x=0$,
$x=1$, $x=2$, and $x=3$.  
Parameters are $\Rey = 250$, $\rho = 0.086$, $L=32$, $\lambda = 4.83$.
At $x=0$, the velocity is
reflection-symmetric in $y$ and primarily spanwise.  The ribbon is
seen as the area $|y| < \rho = 0.086$ with no flow.  Streamwise
vortices are visible for $x \geq 1$.  The scale for distances in $x$
is stretched by a factor of 3.333 relative to distances in $y$,$z$.
}
\label{fig:evecx}
\end{figure}

\begin{figure}[ht]
\vspace*{-8cm}
\centerline{\psfig{file=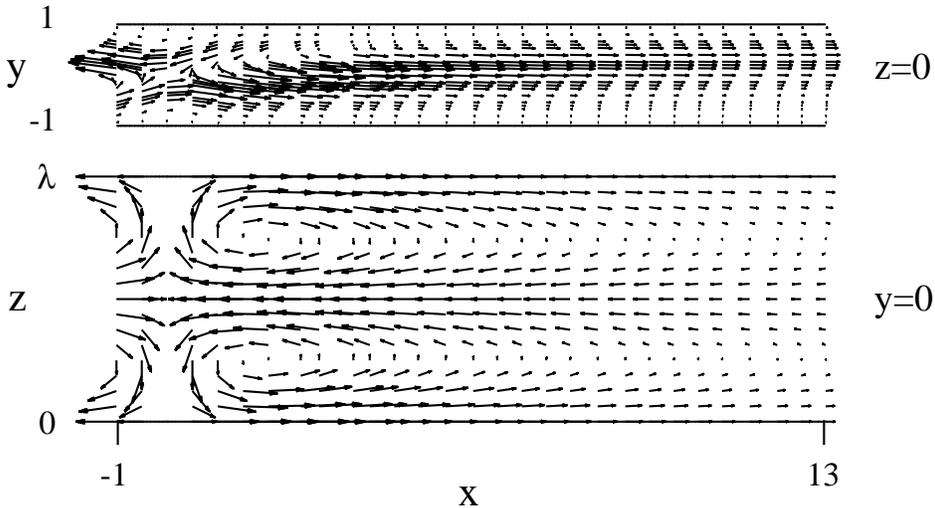,width=17cm}}
\vspace*{-6cm}
\caption{{}Velocity field of near-marginal eigenvector
in the planes $z=0$ and $y=0$.
}
\label{fig:evecyz}
\end{figure}

A computed eigenvector $\emode = (\tilde{u},\tilde{v},\tilde{w})$ 
is shown in Figs.~\ref{fig:evecx} and \ref{fig:evecyz}. 
This eigenvector is near-marginal: $\Rey = 250$, close to $\Rey_c = 230$.
The spanwise wavelength is $\lambda = \lambda_c \equiv 2 \pi/\beta_c =
4.83$.  
The other parameters are $\rho = 0.086$ and $L=32$.

Figure \ref{fig:evecx} shows $(\tilde{v},\tilde{w})$ velocity plots at
four streamwise locations.  In the $x=0$ plane containing the ribbon,
the flow is reflection-symmetric in $y$, and the flow is primarily
spanwise.  The trigonometric dependence in $z$ with the choice of
phase (\ref{bmode}) can be seen.  In the planes $x=1$, $x=2$, and
$x=3$, two counter-rotating streamwise vortices are present.  The flow
for negative $x$ is obtained by reflection in $y$.

\begin{figure}[ht]
\vspace*{-4cm}
\centerline{\psfig{file=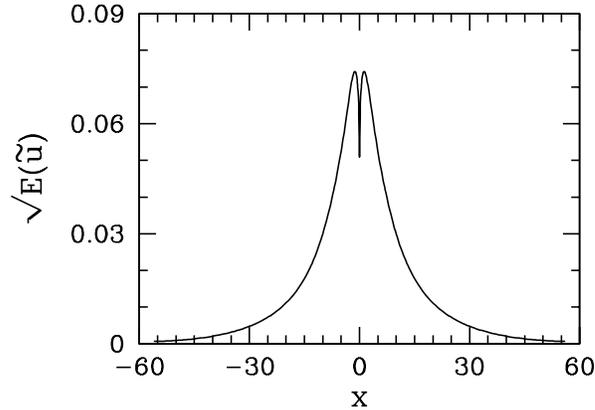,width=12cm}}
\vspace*{-1.5cm}
\caption{{} Energy of near-marginal eigenvector as a function of $x$.
Parameters are the same as in Figs.~\protect{\ref{fig:evecx}}
and \protect{\ref{fig:evecyz}} except that here $L=56$.
Vertical scale is arbitrary.}
\label{fig:envec}
\end{figure}

Figure \ref{fig:evecyz} presents two complementary views of the
eigenvector $\emode$ for $-1 < x < 13$.  Above is a plot of
$(\tilde{u},\tilde{v})$ in the plane $z=0$ where they are maximal
(cf. eq. \ref{bmode}).  Below is a plot of $(\tilde{u},\tilde{w})$ in
the plane $y=0$ at mid-channel height.
Figure \ref{fig:envec} shows the 
$x-$dependence of the spanwise-averaged energy per unit length
\begin{equation}
E(\emode) \equiv {1\over {\lambda_c}} \int_0^{\lambda_c}
dz\int_{-1}^{+1} dy\;{1\over 2}|\emode|^2 
\label{eq:endef}
\end{equation}
Here, the eigenvector was computed in a larger
domain ($L=56$) in order to determine its long-range behavior.
The eigenvector is localized: the energy decays exponentially
with $|x|$ and does not reflect the counter- and co- rotating regions
of the 2D base flow seen on Fig.~\ref{fig:enbmc}.
The flow deficit due to the ribbon produces the local minimum at $x=0$.

We have also computed the vorticity of the eigenvector.  Despite the
streamwise vortices visible on Fig. \ref{fig:evecx}, $\omega_x$ is
the smallest vorticity component and $\omega_z$ by far the largest
over most of the domain.

\subsection{3D Nonlinear stability results}
\label{sec:nonlin} 

Our method of nonlinear stability analysis has previously been used 
to determine the nature of the bifurcation to three-dimensionality
in the cylinder wake \cite{Henderson96}.
The method is based on tracking the
nonlinear evolution of the 3D flow starting from an initial condition
near the bifurcation at $\Reycrit$.  ``Near'' refers both to phase
space (\ie a small 3D perturbation from the two-dimensional profile)
and to parameter space (\ie at a Reynolds number slightly above the
linear instability threshold).  In essence we follow the dynamics
along the unstable manifold of the $2D$ steady flow far enough to 
determine how the nonlinear behavior deviates from linear evolution.  
From this we can determine very simply whether the instability is 
subcritical or supercritical.

Three-dimensional simulations are carried out 
for $\rho = 0.086$ 
at $\Rey=250$, slightly
above $\Reycrit=230$, starting with an initial condition of the form:
\begin{equation}
{\bf u}(x,y,z) = \base(x,y) + \epsilon \emode(x,y,z),
\label{eq:ic}
\end{equation}
where $\base$ is the 2D base flow at $\Rey=250$, $\emode$ is its
eigenmode at wavenumber $\beta_c=1.3$, and $\epsilon$ is a small
number controlling the size of the initial perturbation.

The restriction to wavenumbers which are multiples of $\beta_c$
accurately captures the evolution from initial condition
(\ref{eq:ic}), since the Navier--Stokes equations preserve this
subspace of 3D solutions.  That is, we seek only to follow the
evolution in the invariant subspace containing the critical eigenmode.
We do not address the issue of whether the $\lambda_c$-periodic flow
is itself unstable to long-wavelength perturbations.

To analyze the nonlinear evolution, we define the (real) amplitude $A$ 
of the 3D flow as:
\begin{equation} 
[A \equiv\left[
{1\over {\lambda_c}} \int_0^{\lambda_c} dz
\int_{-1}^{+1} dy\; \int_{-L}^{+L} dx \; {1\over 2}|{\bf u}_1|^2 
\right]^{1/2},
\label{eqn:Adef} 
\end{equation} 
where ${\bf u}_1(x,y,z,t)$ is the component of the 3D velocity
field at wavenumber $\beta_c$, i.e. $A$ is the square root of the 
energy of the flow at wavenumber $\beta_c$.
(A complex amplitude, not required here, 
would include the phase of the solution in the spanwise direction.)

\begin{figure}[ht]
\vspace*{-3cm}
\centerline{\psfig{file=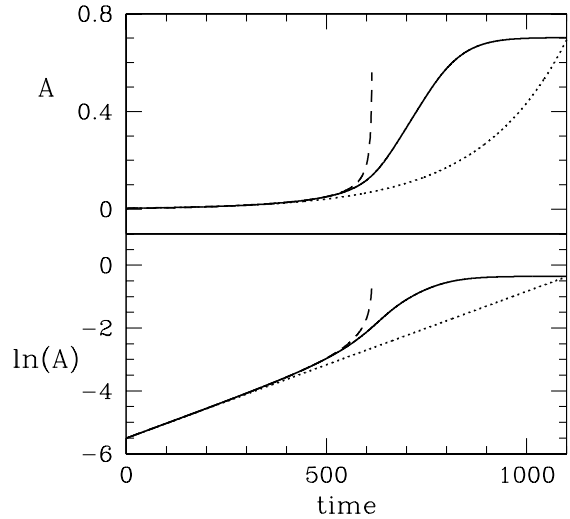,width=12cm}}
\vspace*{-1cm}
\caption{{} Nonlinear growth of the amplitude $A$ of the 3D flow from
simulation (solid) at $\Rey = 250$ plotted on linear and logarithmic
scales.  First-order (dotted) and third-order (dashed) dynamics are
shown with $\sigma=0.004669$ and $\alpha = 0.9$.  The faster than
exponential nonlinear growth (\ie positive $\alpha$) shows that the
instability at $\Rey_c$ is subcritical. }
\label{fig:nonlin}
\end{figure}

Figure~\ref{fig:nonlin} shows the time evolution of $A$ from our
simulations. The value of $\epsilon$ is such that the
initial energy of the 3D perturbation, $\epsilon\emode$, is
$E=A^2=1.66\times10^{-5}$; the energy of the base flow $\base$ is
$E=21.3$.  

To interpret the nonlinear evolution, consider the normal form for a
pitchfork bifurcation including terms up to third-order in the
amplitude:
\begin{equation}
\dot A = \sigma A + \alpha A^3
\label{eq:nf}
\end{equation}
The leading nonlinear term is cubic because the 3D bifurcation is of
pitchfork type (an $O(2)$ symmetric pitchfork bifurcation).  The Landau
coefficient $\alpha$ determines the nonlinear character of the
bifurcation. If $\alpha>0$, then the nonlinearity is destabilizing at 
lowest order and the bifurcation is subcritical; if $\alpha<0$, then 
the cubic term saturates the instability and the bifurcation is supercritical.

Figure~\ref{fig:nonlin} includes curves for first-order evolution (\ie
$\dot A=\sigma A$) and the third-order evolution given by
Eq.~(\ref{eq:nf}).  For the first-order evolution, the eigenvalue
$\sigma$ for the bifurcation has been computed via the linear
stability analysis in Sec.~\ref{sec:linear}.  For the third-order
evolution we have simply fit the one remaining parameter, $\alpha$, in
the normal form.  We followed the approach in
\cite{Henderson96} of using the time series $A(t)$ and the known value
of $\sigma$ to estimate $\alpha$ from $\alpha \simeq (\dot A - \sigma
A)/A^3$.  
This gives $\alpha=0.9\pm0.05$, a constant value for $T \leq 500$,
which determines how long the third-order truncation is valid 
in this case.
The value of $\alpha$ is essentially unchanged
when the mesh is refined by increasing the polynomial order $N$ to 10
or the number $M$ of Fourier modes to 32.  
The magnitude of $\alpha$ depends on the definition of $A$,
but its sign does not. The fact that $\alpha$ is positive indicates that
the instability is subcritical.  Figure \ref{fig:nonlin} indicates
that the 3D flow has become steady by $t \approx 1000$. The nonlinear
saturation seen in the time series is not captured by including a
fifth-order term in the normal form.

\begin{figure}
\vspace{-7cm}
\centerline{\psfig{file=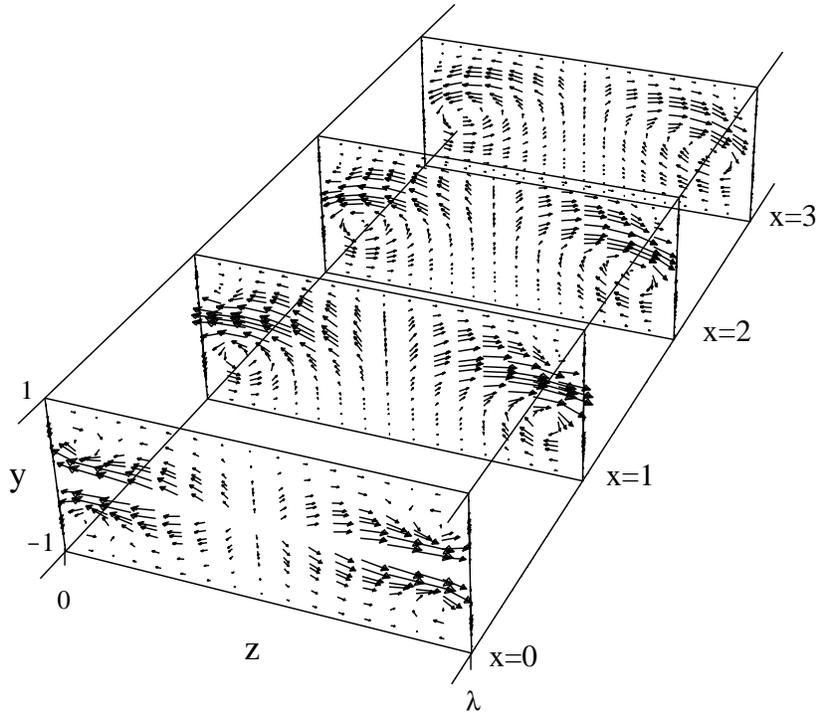,width=12cm}}
\vspace{-3cm}
\caption{{}3D velocity field in the planes $x=0$, $x=1$,
$x=2$, and $x=3$.  
Parameters are $\Rey = 200, \rho = 0.086, L=32, \lambda = 4.83$.
At $x=0$, four small streamwise vortices can be
seen in the corners of the domain.  The lower ($y<0$) vortex pair
evolves into the large vortices seen at $x=1$.  The scale for
distances in $x$ is stretched by a factor of 3.333 relative to
distances in $y$,$z$.
}
\label{fig:R250x}
\end{figure}

We have verified that the instability is subcritical by
computing nonlinear states below $\Rey_c$.
In Fig.~\ref{fig:R250x}, we show the steady 3D flow at $\Rey=200$.
This figure is analogous to
fig. \ref{fig:evecx} depicting the eigenvector, so we will emphasize
here the ways in which the two flows differ.  Small streamwise
vortices can be seen in each of the four corners of the $x=0$ plane
containing the ribbon.  The lower ($y < 0$) pair evolve with $x$ into
the strong pair of vortices at $x=1$.  The vortices at $x=3$ are
tilted with respect to their counterparts in the eigenvector,
attesting to the nonlinear generation of the second spanwise harmonic
$2 \beta$.  The 3D flow in the $y=0$ and $z=0$ planes (after
subtraction of the dominant 2D base flow) is sufficiently similar to
the eigenvector (fig. \ref{fig:evecyz}) that we do not present it here.

\begin{figure}
\vspace{-4cm}
\centerline{\psfig{file=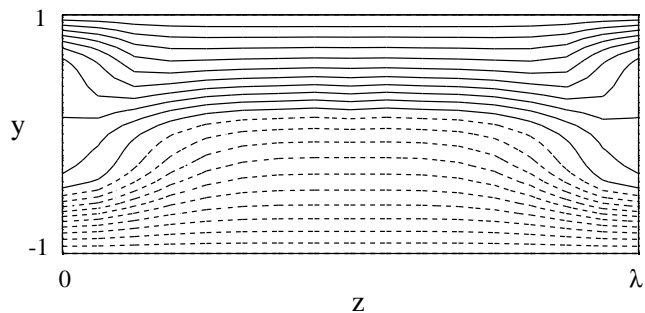,width=10cm}}
\vspace{-3cm}
\caption{{}Streamwise velocity contours in the plane $x=2$
for the 3D field. 
Solid contours correspond to $u > 0$, dashed contours to $u < 0$.
$u$ is small and nearly constant in the interior near $z=0,\lambda$; 
elsewhere it varies approximately linearly with $y$.
}
\label{fig:xslice}
\end{figure}

Streamwise velocity $u$-contours of the 3D flow 
at $x=2$ are shown in Fig. \ref{fig:xslice}.
The $(v,w)$ projections of our 3D flow in Fig.~\ref{fig:R250x},
showing the tilted streamwise vortices, resemble the depictions
of optimally growing modes by Butler and Farrell \cite{Butler}, 
of instantaneous turbulent flows by Hamilton et al. \cite{Hamilton} 
and of weakly forced states by Coughlin \cite{Coughlin}.
However, our streamwise velocity $u$ pictured in 
Fig.~\ref{fig:xslice} differs significantly
from \cite{Hamilton,Coughlin} in that their $u$ contours
are much more strongly displaced at the vortex boundaries.
This is probably due to the fact that our Reynolds number
of 200 is substantially lower than their $\Rey=400$.

\begin{figure}
\vspace{-4cm}
\centerline{\psfig{file=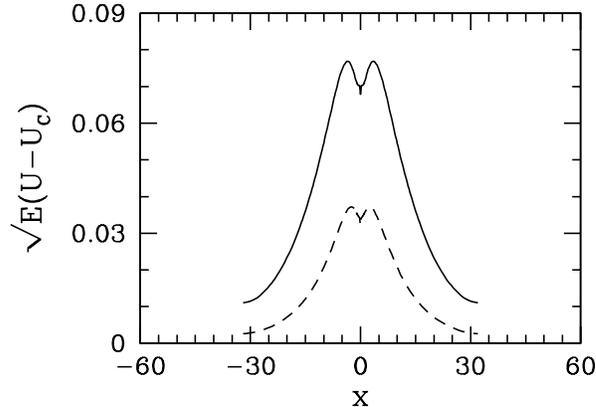,width=12cm}}
\vspace{-1cm}
\caption{{} Energy of
deviation from plane Couette flow of 3D velocity fields at $\Rey = 200$
(dashed) and $\Rey = 250$ (solid) as a function of $x$.
Plotted is the square root of energy per unit length in $x$.
The range in $x$ is taken larger than the computational domain ($L=32$) to
match the range of 
Figs.~\ref{fig:enbmc} and \ref{fig:envec}.}
\label{fig:en3dim}
\end{figure}

Far from the ribbon, the 3D flow returns to plane Couette flow.
Fig.~\ref{fig:en3dim} compares the energy distribution $E(\base-\UC)$ 
defined by (\ref{eq:endef}) of the
deviation of the 3D flow from plane Couette flow
at $\Rey = 200$ with that at $\Rey = 250$.
Note that the 3D flow is less localized than
the corresponding eigenvector (Fig.~\ref{fig:envec}).  
It can be seen that at the higher Reynolds
number the deviation has higher energy, and importantly, occupies a
larger streamwise extent.  This is in accord with the experimental
observation that the streamwise extent of vortices in the perturbed
flow increases with increasing Reynolds number.

We have attempted to determine the location of the saddle-node bifurcation
marking the lower Reynolds number limit of this branch of steady 3D states;
we believe that it occurs just below $\Rey=200$.
There remains nevertheless a slight uncertainty regarding the lower
bound for these states because we have found evidence of two different types
of branches of steady 3D states over the range $200 \leq \Rey \leq 250$.
The study of these states is further complicated by
the fact that the time evolution to many of them
is oscillatory, indicating that their least stable eigenvalues
are a complex conjugate pair.
Further investigation is required
to ascertain the full nonlinear bifurcation diagram.

We have also sought to determine how the scenario
changes as the ribbon radius $\rho$ is decreased. 
Recall from section \ref{sec:linear} that 
for $\rho = 0.043$, we found $Re_c \approx 550$. 
At these parameter values, 3D simulations display
chaotic time evolution.
By decreasing $\Rey$, we have succeeded in computing 
a stable 3D steady state at $\Rey=350$. Since the 
simulation showed chaotic oscillation for a long time
(3000 time units) before showing signs of approaching
a steady state, there remains the possibility that
stable 3D steady states are also attainable for higher $\Rey$.
Simulations at $\Rey = 300$ result in decay to the basic 2D state
(although we do not exclude the possibility of maintaining 3D states 
by a more gradual decrease in $\Rey$).
This is consistent with simulations of the unperturbed
plane Couette geometry ($\rho = 0$) by Hamilton et al. \cite{Hamilton},
who observed chaotic oscillation for $\Rey \geq 400$ and
plane Couette flow $\UC = y{\bf \hat x}$ for $\Rey = 300$.
Other numerical \cite{Lundbladh} and experimental 
\cite{Tillmark,Daviaud} investigations in the unperturbed
plane Couette geometry also indicate a critical Reynolds
number of 360-375 for transition to turbulence.
We plan to investigate the $\rho$-dependence of
the steady 3D states and their stability in a future publication.

\section{CONCLUSION}
\label{sec:discuss}

We have performed a computational linear and nonlinear stability
analysis of perturbed plane Couette flow in order to understand 
experiments recently performed at Saclay
\cite{Dauchot,Bottin,Manneville}, and more generally,
three-dimensional flows in the plane Couette system.
We have accurately determined the extent to which the basic steady 2D
profile is modified by the presence of a small spanwise-oriented 
ribbon in the flow.  
We have determined that such a ribbon, comparable in size to the cylinders
used in the Saclay experiments, is large enough to induce linear
instability of the basic profile at Reynolds numbers of order a few
hundred.  

We elaborate further on how our analysis complements the Saclay results. 
An experimental diagram was obtained \cite{Bottin,Manneville}
for the Reynolds number range of existence of various types of flows:
2D, 3D with streamwise vortices, intermittent, and turbulent.  
In these experiments it was not determined
whether the 3D streamwise vortices arise from a linear instability of the
2D flow.
Our results show that a small geometric perturbation does destabilize
the 2D flow in a subcritical instability 
and that the bifurcating solution is a 3D flow with
streamwise vortices.
Specifically, 
for a nondimensional radius $\rho = 0.086$, we find $\Rey_c = 230$ 
and for $\rho = 0.043$, we find $\Rey_c = 550$.
The computed spanwise wavelength of the most
unstable mode is in good agreement with the value seen
experimentally.  
The streamwise extent occupied by these vortices
decreases with decreasing Reynolds number, 
as observed in experiment, 
and is finite at the lower Reynolds limit of the 3D flows.

The Reynolds number ranges for the steady 3D flows 
we have computed differ somewhat from those seen experimentally.
For $\rho = 0.086$, 
streamwise vortices were observed experimentally over 
the range $150 \lesssim \Rey \lesssim 290$.
For $\rho = 0.086$, we have thus far found steady 3D flows only 
if $\Rey \geq 200$.
In experiments with $\rho = 0.043$, streamwise vortices have been
observed over the range $190 \lesssim \Rey \lesssim 310$;
we have thus far found steady 3D flows only for $\Rey$ near 350.
However, a full study of the 3D flows is still pending and may resolve 
these discrepancies. 

There are also minor qualitative differences between our results 
and the experimental findings. The first is that 2D
flows in our computations are more antisymmetric in the cross-channel
direction $y$ than in experiment (our Fig.~\ref{fig:base_profiles}
{\it vs.} Fig.~4 of \cite{Manneville}).  This is probably due to the
fact that we perturb our flow with an infinitely thin ribbon and not a
wire (cylinder) as in the experiment.  However, based on the existence
of instability in both cases, this small difference in the basic 2D
flow is probably not very significant.  The other difference
is due to the fact that our 3D simulations
impose spanwise periodicity with a single critical wavelength
$\lambda_c$. Experimentally, it is observed that the streamwise
vortices are not always regularly spaced in the spanwise direction.

As stated in the introduction, streamwise vortices can be made to
appear in channel flows via a number of approaches.  Butler and
Farrell \cite{Butler} and Reddy and Henningson \cite{Reddy} show that 
under linear evolution modes of this type can achieve very high amplitude 
before eventually being damped.  In the pseudospectrum interpretation of
Trefethen et al. \cite{Trefethen,Reddy}, non-normality leads to sensitivity 
of the spectrum: streamwise vortices are unstable modes of a slightly
perturbed linear stability matrix.  Our results are entirely
consistent with this interpretation: the ribbon or wire serves as a
specific realization of a perturbation to the stability matrix, and
has indeed rendered the flow linearly unstable to streamwise vortices.

Future computational work is needed to explore these flows.  
Calculating complete bifurcation diagrams for the two cases
$\rho = 0.086$ and $\rho = 0.043$ is the first priority.
We plan to study quantitatively and qualitatively the bifurcations
at which the steady 3D flows terminate at low $\Rey$ and
lose stability at high $\Rey$.
Our goal is to continue 2D and 3D solutions of the perturbed system
to the plane Couette case, $\rho=0$. 

\acknowledgments
This research was in part conducted while the authors were
visitors at the Institute for Mathematics and its Applications (IMA)
of the University of Minnesota, which is supported by the National
Science Foundation.
Some of the results were obtained using the supercomputer 
facilities of the Institut du D\'eveloppement et des Ressources
en Informatique Scientifique (IDRIS).
We thank S. Bottin, A. Cherhabili, O. Dauchot, and K. Coughlin for 
discussing their results with us prior to publication.
We also thank U. Ehrenstein, J. Guckenheimer, P. Manneville, 
M. Rossi, and L.~N. Trefethen for interesting discussions.
We gratefully acknowledge R.D. Henderson for the use of {\em Prism}.

\bibliography{pf5}
\end{document}